\documentclass[preprintnumbers,amsmath,amssymb]{revtex4}

\usepackage{amsmath}
\usepackage{amssymb}
\usepackage{graphicx}
\usepackage{dcolumn}
\usepackage{amsfonts}
\usepackage{amscd}

\newtheorem{theorem}{Theorem}[section]
\newtheorem{lemma}{Lemma}[section]

\newtheorem{definition}{Definition}[section]
\newtheorem{proof}{Proof}[section]

\newcommand{\lv}{\left \vert}
\newcommand{\rv}{\right \vert}
\newcommand{\la}{\left \langle}
\newcommand{\ra}{\right \rangle}
\newcommand{\ket}[1]{\lv #1 \ra}
\newcommand{\bra}[1]{\la #1 \rv}

\begin{document}

\markboth{Yu Tanaka, Damian Markham, Mio Murao} {Local encoding of
classical information onto quantum states}

\title{Local encoding of classical information onto quantum states}

\author{Yu Tanaka$^1$, Damian Markham$^1$, Mio Murao$^{1,2}$
\\\vspace{6pt}
{\it $^1$Department of Physics, Graduate School of Science,
  University of Tokyo, Tokyo 113-0033
Japan\\
$^2$PRESTO, JST, Kawaguchi, Saitama 332-0012, Japan}
\vspace{6pt}\received{v2.0}}

\begin{abstract}
In this article we investigate the possibility of encoding classical information onto multipartite
quantum states in the distant laboratory framework. We show that for all states generated by Clifford
operations there always exists such an encoding, this includes all stabilizer states such as cluster
states and all graph states. We also show local encoding for classes of symmetric states (which cannot
be generated by Clifford operations). We generalise our approach using group theoretic methods
introducing the unifying notion of {\it Pseudo Clifford} operations. All states generated by Pseudo
Clifford operations are locally encodable (unifying all our examples), and we give a general method for
generating sets of many such locally encodable states.
\end{abstract}

\maketitle

\section{Introduction}

In quantum information we are often operating in a framework of
separate distant laboratories, and the questions of what is possible
or impossible under these local restrictions is crucial for
understanding how we can use quantum resources in the best way.
Through these considerations we have come to see entanglement as a
resource for example for quantum cryptography
\cite{Bennett84,Ekert91}, teleportation \cite{Bennett93}, dense
coding \cite{Bennett92} and measurement based quantum computing
\cite{Raussendorf01}. Beyond this however, when considering local
access of information encoded onto quantum states, we have also seen
non-local features without the presence of entanglement
\cite{Bennett99}.

We now turn to consider the complement of this problem, that of
local encoding of information. As well as being interesting in its
own right as a local restricted task, and as a complement to local
access of information (and associated notions of locality), the
ability to locally encode information is in fact an important part
of many quantum information protocols. Indeed the first stage of
dense coding (in fact, the encoding part) is precisely local
encoding of classical information. It is also strongly connected to
the problem of local unitary equivalence which plays a large role in
entanglement theory (as we will see below).

In this paper we raise the question ``can we locally encode on all
states?''. That is, given an arbitrary state, can we use this as a
quantum resource and encode the maximum classical information
possible. Although this is a simple question, the answer is
surprisingly difficult to find, and we find that we cannot give a
positive or negative answer, other than to give a large set of
examples where we can encode, and show how to encode and we are
unable to find any example that we cannot locally encode.

Since classical information is completely distinguishable, the problem of encoding classical
information on a state, becomes the problem of generating an orthogonal bases from that state. Then,
our question becomes ``is it possible to locally generate a complete basis from all states?''. In this
way local encoding is related to local unitary equivalence of states. Though we may naively expect this
to be simple to answer, there are hints that it is a hard question, explaining why we have been unable
to find the solution. The strongest such hint comes from the existence of unextendible product bases
\cite{Bennett99b}. This is a set of orthogonal product states who's compliment must be entangled - that
is, it is impossible to find another state orthogonal this set which is product. This is an example
where if we do not choose appropriate encoding operators, full local encoding is not possible (even
though there may exist another set of local operations to encode full classical information).

We begin in section \ref{SCN: General} by considering the local
encoding of general product states. Our approach is then to take
this encoding and extend it to sets of entangled states generated by
unitaries which obey certain commutation relations. We concentrate
on using Pauli operations to encode on the states, where we develop
the notion of {\it Pseudo} Clifford operations whose properties
allow us to give a general sufficient condition for the ability to
locally encode on a state. In particular this gives a method for
locally encoding on all stabiliser states, including cluster states
used in measurement based quantum computation \cite{Raussendorf01},
CSS error correction code states and all graph states \cite{Hein06}.
In section \ref{SCN: symm states} we show local encoding for sets of
symmetric states as examples of non-Clifford but Pseudo Clifford
states.  In section \ref{SCN: Group analysis} we use group analysis
to investigate what states can be locally encoded by the methods we
have introduced. We note that although in our methods we are
restrictive on the allowed encoding operations (i.e. Pauli and
derived from the product state case), our approach manages to cover
all the states we consider here, and we have no example of states
which we can show cannot be encoded by our methods.

\section{Local encoding on Pseudo Clifford states}
\label{SCN: General}

The problem of local encoding of classical information is equivalent
to that of generating a basis by local operations. We begin by
giving a formal definition of local encoding:
\begin{definition}
\label{def:localencoding} A $n$-qubit quantum state $\ket{\psi}$ is said to be {\it locally encodable}
if there exists a set $\{ v_i\ |\ v_i \in SU(2)^{\otimes n} \}_{i=0}^{2^n-1}$ of local unitary
operations such that $\bra{\psi}v_i^{\dagger}v_j\ket{\psi} = \delta_{ij}$ for all $i,j$. We call such $
v_i $ {\it local encoders}, and the set $ \{ v_i \} $ the {\it local encoder set}.
\end{definition}
By this definition, we can always ignore the global phase of the
encoded state $v_i \ket{\psi}$, since we only require orthogonality
between the encoded states.

The difficulty of finding the local encoder set lies in that the
condition $\bra{\psi}v_i^{\dagger}v_j\ket{\psi} = \delta_{ij},
\forall i,j$ in Definition \ref{def:localencoding} is a weak
condition, it only restricts the property of a local encoder set $\{
v_i \}$ applied to the given state $\ket{\psi}$. Given a state
$\ket{\psi}$ it is difficult to check if there exists an encoding
set, since we must search over all possible local unitaries. In this
sense it is too weak to be easily checkable.

In this paper, we take another approach to understand the properties of local encoding; First we pose a
restriction on the local encoder to be tensor products of the Pauli operations $\{I, X, Y, Z \}$
represented by
\begin{eqnarray}
 X =\left(
  \begin{array}{cc}
      0 & 1   \\
      1 & 0   \\
  \end{array}
\right), Y=\left(
  \begin{array}{cc}
      0 & -i   \\
      i & 0   \\
  \end{array}
\right), Z=\left(
  \begin{array}{cc}
      1 & 0   \\
      0 & -1   \\
  \end{array}
\right),
\end{eqnarray}
in the computational basis $\{\ket{0},\ket{1} \}$ and obtain
sufficient conditions for locally encodable states.  Then we
consider gradually relaxing the restriction to find wider classes of
states which are locally encodable and construction of the local
encoder sets.

Our restriction of the local encoder set to consist of the Pauli operations allows group theoretical
analysis. The set of $4^n$ $n$-tensor product operations consisting of the Pauli operators $\{I, X, Y,
Z \}$, together with their overall phase of $\pm 1$ or $\pm i$ forms a group, which is called the {\it
Pauli group}, denoted by $\mathcal{P}$ (here we say the Pauli group on $n$-qubits is given by the
tensor products of all qubit Pauli operators). Including the phase factor, the number of elements of
the $n$-qubit Pauli group is $4 \cdot 4^n$.  For local encoding, we ignore the global phase factor and
only care about $4^n$ elements of Pauli group operators.

To investigate local encoding, we use the properties of the Pauli group and another group, the {\it
Clifford group}. The Clifford group is a group consisting of all operators which leave $\mathcal{P}$
fixed under conjugation, and
 is denoted by $\mathcal{C}$. Formally, we write it as the set of operators $\{ C \in SU(2^n) | C p C^{\dagger} \in
\mathcal{P},\ \forall p \in \mathcal{P} \}$.  The Clifford group is
generated by combinations of the Hadamard gate $H$, the Phase gate
$S$ and the control-NOT gate $U_{CNOT}$ represented in the
computational basis by
\begin{eqnarray}
 H =\frac{1}{\sqrt{2}}\left(
  \begin{array}{cc}
      1 & 1   \\
      1 & 1   \\
  \end{array}
\right),~S=\left(
  \begin{array}{cc}
      1 & 0   \\
      0 & i  \\
  \end{array}
\right),~U_{CNOT} = \left( \begin{array}{cccc}
      1 & 0 &0 &0   \\
      0 & 1&0 & 0   \\
      0 & 0&0 & 1   \\
      0 & 0&1 & 0   \\
  \end{array}
\right),
\end{eqnarray}
in addition to the Pauli operations.

\bigskip

We first consider encoding onto product states.  We denote a local
encoder for a zero state, which is $n$-tensor products of zero
states defined by $\ket{\bar{0}} \equiv \ket{0}^{\otimes n} =
\ket{0} \otimes \ket{0} \otimes ...\otimes \ket{0}$, by $\{ w_i \}
$. For a simple product state such as $\ket{\bar{0}}$, it is
apparent that a local encoder set is given by $\{w_0=I \otimes
...\otimes I \otimes I$, $w_1=I \otimes ... I \otimes X$, $w_2=I
\otimes ... X \otimes X$, .... ,$w_{2^n-1}=X \otimes ... X \otimes
X\}$. This construction is based on the fact that we can ``flip''
each single qubit state $\ket{0}$ to $\ket{1}$ by performing a Pauli
operation $X$. All combinations of $\{ I, X\}$ for $n$-qubits acting
on $\ket{\bar{0}}$ give a set of states, which is a complete
orthonormal set of the states denoted by $\{ \ket{\bar{i}}\}$.  For
convenience, we express this local encoder of the zero state by
\begin{eqnarray}
\{w_i^{\bar{0}}\equiv w_i=X^{m_1} \otimes ... \otimes X^{m_n} \},
\label{eqn:zeroencoder}
\end{eqnarray}
where a set of indices $\{ m_1,m_2,...,m_{n} \}$ is a binary
representation of $i$.

We now generalize this way of local encoding to any product state. A
general $n$-qubit product state $\ket{\phi_{prod}}$ is represented
by $n$-tensor products of single qubit states $\ket{\phi_k}=\cos
(\theta_k/2) \ket{0}+e^{i \varphi_k } \sin (\theta_k/2) \ket{1}$
where $\theta_k$ and $\varphi_k$ are positive parameters for $k$th
qubit satisfying $0 \le \theta \le \pi$ and $0 \le \phi \le 2 \pi$.
These parameters represent the angles of the state vector
$\ket{\phi_k}$ on the Bloch sphere. We consider how to perform a
flip operation using a {\it minimum} number of parameters for a
general $\ket{\phi_k}$ to make it as simple and general as possible.
It is known that there is no universal flip operation for arbitrary
states \cite{NielsenChuang}. If we know the two real parameters
$\theta_k$ and $\varphi_k$, we can always transform $\ket{\phi_k}$
back into $\ket{ {\bar 0}}$, so it is trivial to find the flip
operation. In fact, we can describe a flip operation with just a
single real parameter, since a state in the form of $\ket{\phi_X}
\equiv \cos (\theta_k/2) \ket{0}+i \sin (\theta_k/2) \ket{1}$ can be
flipped by $X$ operation irrespective of the parameter $\theta_k$.
In the Bloch sphere picture, the state $\ket{\phi_X}$ is on the
$yz$-plane, therefore the $X$ operation (which is proportional to a
$\pi$-rotation around the $x$-axis) transforms $\ket{\phi_X}$ into
its orthogonal state $\ket{\phi_X^\perp}$.  Noting that a general
single-qubit state $\ket{\phi_k}$ is transformed to $\ket{\phi_{X}}$
by a modified phase gate (a rotation around the $z$-axis until the
state lies on the $z-y$ plane)
\begin{eqnarray}
s(\varphi_k) = \left(
  \begin{array}{cc}
      1 & 0   \\
      0 & i e^{-i \varphi_k}   \\
  \end{array}
\right),
\end{eqnarray}
the flip operation for each $\ket{\phi_k}$ is given by $X \cdot
s(\varphi_k)$. Thus a possible local encoder set of a general
product state $\ket{\phi_{prod}}$ is given by $\{ w_i^{\bar{0}}
\cdot s(\varphi_k)^{\otimes n} \}$ where $s(\varphi_k)^{\otimes n}=s
( \varphi_1 ) \otimes ... \otimes s(\varphi_n)$ is independent of
$i$.

\bigskip

Our strategy now is to try to use our knowledge about the local
encoder set of arbitrary product states for constructing the local
encoders of entangled states. For this purpose, we introduce a
representation of a $n$-qubit state $\ket{\psi}$ by using a
(generally non-local) unitary operation $U$ on the zero state
$\ket{\bar{0}}$, namely, $\ket{\psi} = U \ket{\bar{0}}$. We note
that by this relation, the unitary operation $U$ is not uniquely
determined, only the first column of $U$ is determined (in the
computational basis).

\begin{figure}
\begin{picture}(450,90)
\put(170,00){\includegraphics[scale=1.0]{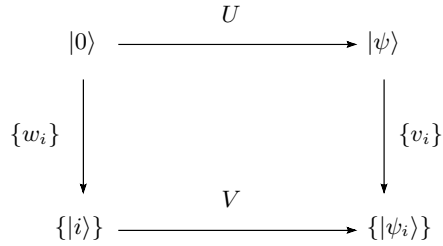}}
\put(167,70){$\ket{0}$} \put(162,0){$\{ \ket{i} \}$}
\put(280,70){$\ket{\psi}$} \put(280,0){$ \{ \ket{\psi_i} \}$}
\put(145,35){$\{ w_i \}$} \put(292,35){$\{ v_i \}$}
\put(225,80){$U$} \put(225,10){$V$}
\end{picture}
\caption{ The relationship between operations $ v_i $, $w_i $, $U$,
and $V$ for local encoding of $\ket{\psi} = U \ket{\bar{0}}$ by the
local encoder set $\{ v_i \}$. The set $\{w_i\}$ encodes $|0\rangle$
to basis $\{|i\rangle\}$, and the set $\{ v_i \}$ locally encodes
$\ket{\psi} = U \ket{\bar{0}}$ to $\{|\psi_i\rangle\}$, in
accordance with Eq.~(\ref{EQN: gen condition II}).}
\label{fig:relationship}
\end{figure}

Using the unitary representation, the conditions for local encoding are written by $\bra{{\bar{0}}}
U^\dagger v_i^{\dagger} v_j U \ket{{\bar{0}}} = \delta_{ij}$. We rewrite this condition as
\begin{eqnarray} \label{EQN: gen condition II}
\bra{{\bar{0}}} U^\dagger v_i^{\dagger}V V^\dagger v_j U
\ket{{\bar{0}}} = \bra{{\bar{0}}} W_i^{\dagger} W_j \ket{{\bar{0}}}
 =\delta_{ij}
\end{eqnarray}
where $V$ is an arbitrary (non-local) unitary operator and $W_i
\equiv V^\dagger v_i U$.  Although the operators of $\{ W_i \}$ are
not necessary to be local in general, we investigate the cases where
they are restricted to be local, therefore, they are given by the
local encoder $\{ W_i=w_i \}$, for our purpose. We can view the
relationship of $v_i$, $w_i$, $U$, and $V$ in Figure
\ref{fig:relationship}.

We now introduce a simplification. We consider the case that {\it
both} $\{v_i\}$ and $\{w_i\}$ are given by $n$-tensor products of
Pauli operations and also impose $V=U$.  Then we can immediately see
that if $U$ is Clifford, i.e. it maps every tensor product of Pauli
operators into another tensor product of Pauli operators, thus if
$\{w_i\}$ are Pauli, then $\{v_i=U w_i U^\dagger \}$ are also Pauli.
We know that a local encoder set of the zero state $\ket{{\bar{0}}}$
is given by $\{{w_i^{\bar{0}}} \}$ of Eq.(\ref{eqn:zeroencoder}),
therefore, all the Clifford group states represented by
$\ket{\psi}=U \ket{\bar 0}$ where $U \in {\cal C}$ are locally
encodable by $\{ v_i=U w_i^{\bar{0}} U^\dagger \}$.

We note that our condition of $ v_j  = U w_i U^\dagger \in \mathcal{P}$ need be only satisfied by the
$2^n$ elements of the local encoder $\{ w_i \}$, out of $4^n$ different tensor products of the Pauli
group operations. Thus, $U$ can be taken to be in a larger class of operators than the Clifford group
operators. We call the set of operators $\{ U \}$ which satisfy $ v_j  = U w_i U^\dagger \in
\mathcal{P}$ (where $\{w_i\}$ are also Pauli) for $2^n$ different tensor products of the Pauli
operators the {\it Pseudo Clifford} set, denoted by ${\cal PC}$ (a formal definition will be given in
section \ref{SCN: Group analysis}). We will present constructions of local encoder sets for states
which are not generated by the Clifford group operations, but the Pseudo Clifford set operations from
$\ket{{\bar{0}}}$ in section \ref{SCN: symm states}. The group-like properties of the Pseudo Clifford
set is investigated further in section \ref{SCN: Group analysis}. By definition, all Pseudo Clifford
generated states can be locally encoded when the $\{w_i\}$ locally encode $\ket{{\bar{0}}}$.

\bigskip

Now we consider how much we can extend the class of $U$ beyond the
Pseudo Clifford sets.  Obviously, if a given entangled state
$\ket{\psi}=U \ket{{\bar 0}}$ is locally encodable by a local
encoder set $\{ v_i \}$, any locally equivalent state of the given
state denoted by $\ket{\psi^\prime}= V_L \ket{\psi}$ where $V_L =
u_1 \otimes ... \otimes u_n$ and $u_k \in SU(2)$ is locally
encodable by $\{ v_i \cdot V_L^\dagger \}$. This indicates the
degree of freedom for the choice of the basis of the Pauli
operations does not effect the ability of local encoding.

To summarize the results obtained in this section, we have the
following theorem:
\begin{theorem} \label{theorem:deomposition} \textbf{Decomposition condition}\\
A state $\ket{\psi}$ of $n$ qubits can be locally encoded if there
exist unitaries $V_L$ and $U_{PC}$ such that $\ket{\psi}=V_L \cdot
U_{PC}\ket{\bar{0}}$, $V_L$ is local, and $U_{PC}$ is Pseudo
Clifford (such that it conjugates a set $\{w_i\}$ encoding
$\ket{\bar{0}}$ to local Pauli operators). The local encoder set is
then given by
\begin{eqnarray}
\{ v_i = U_{PC}  w_i  U_{PC}^\dagger \cdot V_L^\dagger \}.
\end{eqnarray}
\end{theorem}
\noindent A set of such $\{w_i\}$ is given by $\{w_i^{\bar{0}} \}$
of Eq.~(\ref{eqn:zeroencoder}), or similar sets replacing $X$ with
$Y$, and/or replacing $I$ with $Z$.

\begin{figure}
\begin{picture}(520,180)
\put(137,-7){\includegraphics[scale=1.2]{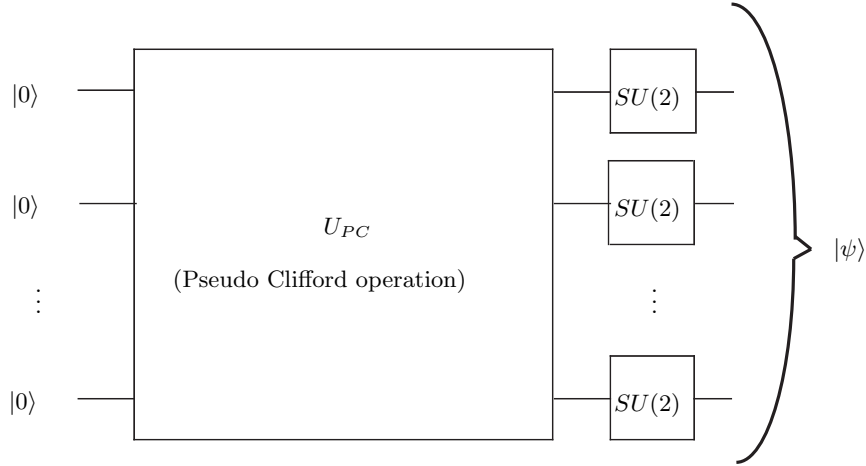}} \put(340,129){$SU(2)$} \put(340,87){$SU(2)$}
\put(340,13){$SU(2)$} \put(230,80){$U_{PC}$} \put(170,60){ \textrm{(Pseudo~Clifford operation)} }
\put(112,130){$\ket{0}$} \put(112,88){$\ket{0}$} \put(111,14){$\ket{0}$} \put(121,50){$\vdots$}
\put(354,50){$\vdots$} \put(423,72){$|\psi\rangle$}
\end{picture}
\caption{All states $|\psi\rangle$ generated by the above circuit
can be locally encoded (c.f. Theorem \ref{theorem:deomposition}).}
\label{fig:decomposotion}
\end{figure}

\begin{figure}
\begin{picture}(520,150)
\put(90,10){\includegraphics[scale=0.9]{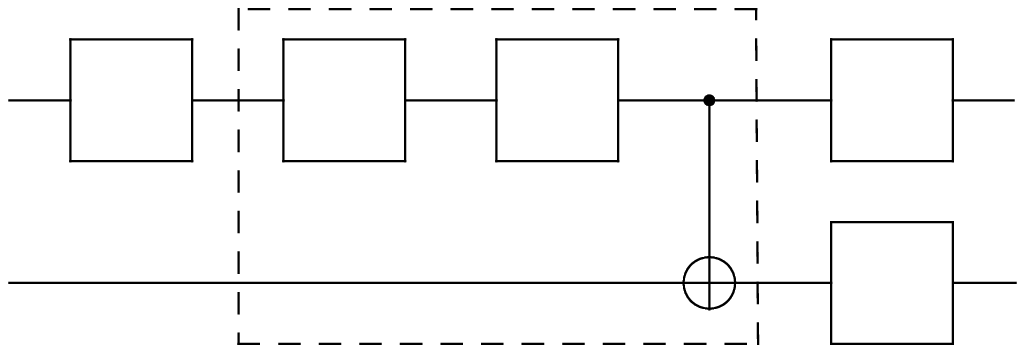}}
\put(174,72){$Z$} \put(316,72){$u_1$} \put(230,72){$S$}
\put(113,72){$r{(\theta)}$} \put(72,71){$\ket{0}$}
\put(72,24){$\ket{0}$} \put(316,24){$u_2$}
\put(199,105){\textrm{Clifford}} \put(360,49){$\Biggr\}$}
\put(370,50){\textrm{Schmidt decomposition of}}
\put(370,30){$\ket{\psi}$}
\end{picture}
\caption{A quantum circuit representing the Schmidt decomposition of
two qubit state $\ket{\psi}=\cos(\theta/2) \ket{a_0} \otimes
\ket{b_0}+\sin(\theta/2) \ket{a_1} \otimes \ket{b_1}$. Operation
$r(\theta)$ is not clifford, but is Pseudo Clifford, hence allowing
local encoding.} \label{fig:twoqubitencoding}
\end{figure}

As an example of this result, we show that any two qubit state is
locally encodable. All two qubit states can be written in the
Schmidt decomposition as $\ket{\psi}=\cos(\theta/2) \ket{a_0}
\otimes \ket{b_0}+\sin(\theta/2) \ket{a_1} \otimes \ket{b_1}$.  The
Schmidt decomposition state can be represented by
\begin{eqnarray}
u_1 \otimes u_2 \cdot U_{CNOT} \cdot (s(\pi) \cdot r (\theta))
\otimes I \ket {\bar 0},
\end{eqnarray}
where $u_1$ and $u_2$ map a computational basis $\{ \ket{i} \}$ into $\{ \ket{a_i} \}$ and $\{
\ket{b_i} \}$, respectively, and
\begin{eqnarray}
r(\theta)= \left(
  \begin{array}{cc}
      \cos(\theta/2) & -i \sin(\theta/2)   \\
      i \sin(\theta/2) & \cos(\theta/2)   \\
  \end{array}
\right),
\end{eqnarray}
 (See
Fig.~{\ref{fig:twoqubitencoding}}). Note that $s(\pi)=S \cdot Z$
therefore it is a Clifford operation. Although $r(\theta)$ is not a
Clifford operation, it is a Pseudo Clifford operation. In fact, it
also commutes with $w_i^{\bar{0}}$, thus this state is always
locally encodable by $\{ v_i= (s(\pi) \otimes I) \cdot U_{CNOT}
\cdot w_i^{\bar{0}} \cdot U_{CNOT} \cdot (s(\pi) \otimes I) \cdot
u_1^\dagger \otimes u_2^\dagger \}$.

Further, Theorem \ref{theorem:deomposition} also means that all
stabiliser states can be locally encoded. This includes cluster
states used in measurement based quantum computation
\cite{Raussendorf01}, CSS error correction code states and all graph
states \cite{Hein06}.

A natural question now is, how large is the class of states covered
by Theorem \ref{theorem:deomposition}. We do not at present have a
good answer, except to say that there are many interesting states
covered, and that it is not trivial to classify in a simple way
those covered. The difficulty arises from the fact that for any
state $\ket{\psi}$, there will be many unitaries $U$ such that
$\ket{\psi}=U\ket{\bar{0}}$ and we only need one to satisfy the
conditions in Theorem \ref{theorem:deomposition} for it to be
locally encodable. A simple example is given by the two qubit
control phase operation (for arbitrary phase) $CP(\theta)$, this is
not in the form of $V_L \cdot U_{PC}$, however as we see above, any
state of two qubits can be written in this form, i.e. there exist a
$V_L$ and $U_{PC}$ such that $CP(\theta)|\bar{0}\rangle=V_L
U_{PC}|\bar{0}\rangle$.

In the remainder of this paper, we will look at explicit sets of examples where Theorem
\ref{theorem:deomposition} can be applied, and use group theory to construct local encoding sets.

\section{Local encoding on symmetric basis states (a large class of examples)}
\label{SCN: symm states}

We now focus on the Pseudo Clifford operation part and investigate
local encoders for the states generated by the non-Clifford but
Pseudo Clifford operations $U \in {\cal PC}$ by imposing an
additional condition $w_i = v_i$ to the definition of the Pseudo
Clifford operation $v_i = U w_i U^\dagger$ for $2^n$ elements of
Pauli operations $\{ w_i \}$. That is, we ask that our encoders
remain the same as for the state $\ket{ \bar{0} }$. This is
satisfied if the $U$ commutes with the $\{w_i\}$, in which case $V$
becomes $U$ and $\{v_i\}\rightarrow\{w_i\}$ in
Fig.~\ref{fig:relationship}, which now is commuting.

Our idea for searching the non-Clifford but Pseudo Clifford operation $U$ is that we investigate
unitary operations which are represented by a sum of several Pauli group operations $\{ p_i \}$,
namely, $U = \sum_i{a_i p_i}$ where $a_i$ is a normalized coefficient ($\sum_i{|a_i|^2}=1$) and the set
$\{ p_i \}$ should be carefully chosen such that the sum represent a unitary operations and we choose
$U$ such that it commutes with our chosen $\{w_i\}$. In this section, we present constructions of the
local encoders for a class of states called {\it symmetric basis states} by finding the sum
representations of the Pseudo Clifford operation generating the states. A general investigation is
given in Section 4.

We first introduce symmetric basis states. Symmetric states are the
states which are invariant under exchange of arbitrary two qubits.
Symmetric basis states are special symmetric states. We define a
$n$-qubit symmetric basis state $\ket{n,m}$ as a symmetric state
consisting of $n-m$ qubits in $\ket{0}$ states and $m$ qubits in
$\ket{1}$ states, namely,
\begin{eqnarray}
\ket{n,m} := \sqrt{\frac{m!(n-m)!}{n!}}\sum_{\pi} \pi \left( \ket{0}
\otimes ... \otimes \ket{0} \otimes \overbrace{\ket{1} \otimes ...
\otimes \ket{1}}^m \right ),
\end{eqnarray}
where $\sum_{\pi}$ is taken by all permutations $\pi$ of the tensor
products of $(n-m)$ $\ket{0}$ states and $m$ $\ket{1}$ states.  A
set of symmetric basis states of $\{ \ket{n,m} \}_{m=0}^n$ forms a
complete basis of the $(n+1)$-dimensional symmetric subspace of the
$n$-qubit Hilbert space ${\cal H}^{\otimes n}$.

We consider the representation of $\ket{n,m}=U \ket{\bar 0}$ using a non-local unitary operation $U$.
The unitary operation $U$ cannot be Pauli operations and Clifford operations for the case of $n >2$ or
the case of product states $m \neq 0$ or $m \neq n$. It requires non-Clifford unitary operators to
generate $\ket{n,m}$ from $\ket{\bar 0}$, since the coefficient of the symmetric basis state
$\sqrt{{m!(n-m)!}/{n!}}$ cannot be obtained by the Clifford group operations which can only give
coefficients of the form $1/\sqrt{2^k}$, where $k =0,1,...,n$. In this section, we show that this $U$
represented by a sum of Pauli group operators becomes the Pseudo Clifford operation for $\ket{n,1}$ and
$\ket{n,n-1}$ symmetric basis states by choosing the appropriate local encoder and the Pauli group
operations in the sum representation of $U$. We extend this method to show that some other symmetric
basis states and related states are also locally encodable.

We first show how to represent $U$ in the Pauli sum representation
for $3$-qubit symmetric basis states $\ket{3,1}$, which is
alternatively called a {\it W state}. We consider the following set
of Pauli operations:
\begin{eqnarray}
\{ v_0 = I \otimes I \otimes I,~
v_1 = X \otimes I \otimes I,~v_2 = Z \otimes  X \otimes I,~v_3 = Z \otimes Z \otimes X, \nonumber \\
v_4 = v_1 v_2,~v_5 = v_2 v_3,~v_6 = v_3 v_1,~v_7 = v_1 v_2 v_3
\}
\label{3encoder}
\end{eqnarray}
Since $v_1$, $v_2$ and $v_3$ forms generators of the above set $\{
v_i \}_{i=0}^7$, we denote the set of operators generated by the
generators by $\langle \{ v_1,\ v_2,\ v_3 \} \rangle=\{ v_i
\}_{i=0}^7$. It is easy to check that the set $\{ v_i \}$ are also
local encoders for $\ket{3,0}=\ket{\bar 0}$ and $\ket{3,3}=X \otimes
X \otimes X \ket{\bar 0}$.

We choose the unitary operation $U_W$ generating the symmetric basis
state $\ket{3,1}=U_W \ket{\bar 0}$ in the Pauli sum representation
as
\begin{eqnarray}
U_W =\frac{1}{\sqrt{3}} (p_0+p_1+p_2) = \frac{1}{\sqrt{3}}(X \otimes
Z \otimes Z+I \otimes X \otimes Z+I \otimes I \otimes X).
\end{eqnarray}
Note that $U_W$ is carefully chosen such that the operators $\{ p_i \}$ in the Pauli sum representation
anti-commute to ensure the unitarity of $U_W$, and that $U_W$ commutes with $v_i$. Therefore,we have
\begin{eqnarray}
\bra{3,1} v_i^{\dagger}v_j \ket{3,1} = \bra{\bar 0}
U_W^{\dagger}v_i^{\dagger}v_j U_W \ket{\bar 0} = \bra{\bar 0}
v_i^{\dagger}v_j \ket{\bar 0} = \delta_{ij},
\end{eqnarray}
and the set of operators $\{ v_i \}_{i=0}^{7}$ given by
Eq.~(\ref{3encoder}) are the local encoder for $\ket{3,1}$.  Since
$\ket{3,2}=X \otimes X \otimes X  \cdot U_W  \ket{\bar 0}$ and $X$
operations applied after $U_W$ do not change the local encoder due
to Theorem~{\ref{theorem:deomposition}}, $\ket{3,2}$ can be also
locally encodable by the same $\{ v_i \}_{i=0}^7$.

By generalizing the construction of the $3$-qubit symmetric basis
states, we show that the $n$-qubit symmetric basis states
$\ket{n,0}$, $\ket{n,1}$ ($n$-qubit W state) $\ket{n,n-1}$ and
$\ket{n,n}$ are also locally encodable.

\begin{theorem}\textbf{Constructive method}\\
Symmetric basis states $\ket{n,0}$, $\ket{n,1}$, $\ket{n,n-1}$ and
$\ket{n,n}$ are locally encodable by the Pauli operators given by
\begin{eqnarray}
\langle \{ g_i = Z^{\otimes (i-1)} \otimes X \otimes I^{\otimes
(n-i)} \}_{i=1}^{n} \rangle. \label{cmencoder}
\end{eqnarray}
\end{theorem}

\begin{proof}
The product state cases ($\ket{n,0}$ and $\ket{n,n}$) are trivial.
We show the proof for the case of $\ket{n,1}$. Defining a unitary
operation in the Pauli sum representation
\begin{eqnarray}
U_{n,1} = \frac{1}{\sqrt{n}}\sum_{i=1}^{n} I^{\otimes (i-1)} \otimes
X \otimes Z^{\otimes (i-1)}, \label{generalW}
\end{eqnarray}
the symmetric basis state $\ket{n,1}$ is represented by $\ket{n,1} =
U_{n,1} \ket{\bar 0}$.  This $U_{n,1}$ satisfies commutation
relation $[U_{n,1},v_i]=0$ for all $i$. From the relationship
\begin{eqnarray}
\bra{n,1}v_i^{\dagger}v_j\ket{n,1} = \bra{\bar 0} U_{n,1}^{\dagger}
v_i^{\dagger} v_j U_{n,1} \ket{\bar 0} = \bra{\bar 0} v_i^{\dagger}
v_j \ket{\bar 0} = \delta_{ij},
\end{eqnarray}
the set $\langle \{ g_i = Z^{\otimes (i-1)} \otimes X \otimes
I^{\otimes (n-i)} \} \rangle_{i=1}^{n}$ are the local encoder of
$\ket{n,1}$. For $\ket{n,n-1}$, it is also locally encodable by the
same local encoder of $\ket{n,1}$, due to the relationship
$\ket{n,n-1}=U_{n,1} \cdot X \otimes X \otimes X \ket{\bar 0}$ and
Theorem \ref{theorem:deomposition}.
\end{proof}

In fact this encoding was used to construct a basis of W states in \cite{Miyake05} in the context of
multipartite entanglement distillation. Here we see if holds for other states as well.

We can also extend the locally encodable states beyond $\ket{n,1}$ and $\ket{n,n-1}$ states using the
constructive method above. By replacing $U_{n,1}$ by $U_{n,1}^\Xi = \sum_{i=1}^{n} a_\pi I^{\otimes
(i-1)} \otimes X \otimes Z^{\otimes(n-i)},\ (a_\pi \in\mathbf{R})$ where $\sum_{\pi} a_\pi^2 =1$, a
state with non-even real weights on permutations can be written in the Pauli sum representation
$\ket{\Xi_{n,1}} = U_{n,1}^\Xi \ket{\bar 0}$. The state $\ket{\Xi_{n,1}}$ is also locally encodable by
the local encoders of $\ket{n,1}$, since our construction does not depend on the coefficient $a_\pi$.

Next, we try to find the local encoders of other symmetric basis
states by using induction.
\begin{lemma} \textbf{Inductive method}\\
If $\ket{n,k}$ and $\ket{n,k-1}$ are locally encodable by the same
local encoder set $\{ v_i \}_{i=1}^{2^n}$, then $\ket{n+1,k}$ is
locally encodable by a new local encoder set $\{ I \otimes v_i, Z
\otimes v_i \}$.
\end{lemma}

\begin{proof}
Since we have $|n,m\rangle = \frac{1}{2}(|0\rangle |n-1,m\rangle
+|1\rangle |n-1,m-1\rangle)$, we see that
\begin{eqnarray}
\bra{n+1,k} Z^i \otimes v_j^{\dagger} I \otimes v_k \ket{n+1,k} &=&
\frac{1}{2}\bigl( \bra{0}Z^i\ket{0} \bra{n,k} v_j^{\dagger}v_k
\ket{n,k}
+ \bra{1}Z^i\ket{1} \bra{n,k-1} v_j^{\dagger}v_k \ket{n,k-1} \bigr)\nonumber \\
&=& \delta_{0i}\delta_{jk}.
\end{eqnarray}
Hence, two states encoded by any two different encoding operators
taken from the set $\{ I \otimes v_i, Z \otimes v_i \}_{i=1}^{2^n}$
are orthogonal. Thus, $\ket{n+1,k}$ is locally encoded by the set
$\{ I \otimes v_i, Z \otimes v_i \}$.
\end{proof}

\begin{figure}

\begin{picture}(470,90)
\put(130,-10){\includegraphics[scale=1.0]{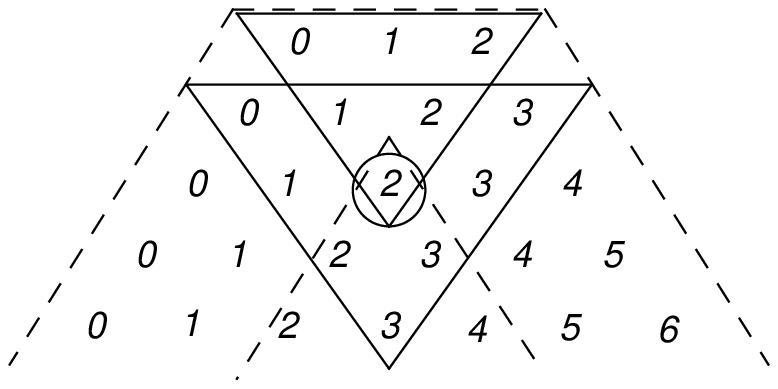}}
\put(100,83){$\ket{2,m}$} \put(100,63){$\ket{3,m}$}
\put(100,43){$\ket{4,m}$} \put(100,23){$\ket{5,m}$}
\put(100,3){$\ket{6,m}$}
\end{picture}

\caption{Local encoding of symmetric basis states.  The number $m$
represents the number of $\ket{1}$ state. The states inside the
dashed line encoded by the construction method by $\langle v_i =
Z^{\otimes (i-1)} \otimes X \otimes I^{\otimes (n-i)}
\rangle_{i=1}^{n}$, the states in inside the triangle line is
encoded by the inductive method by $\{ I \otimes v_i, Z \otimes v_i
\}$.  The state inside the circle is encoded by the local encoder
given by Eq.~(\ref{encoder})} \label{dir}
\end{figure}

All the $3$-qubit symmetric basis states can be locally encoded by
the same local encoders given by Eq.~(\ref{3encoder}), we can see
that the symmetric basis states of $\ket{4,1}$, $\ket{4,2}$,
$\ket{4,3}$, $\ket{5,2}$, $\ket{5,3}$ and $\ket{6,3}$ are locally
encodable from this lemma of the inductive method. For $\ket{4,2}$,
we find that there is another local encoder given by
\begin{eqnarray}
\{ I \otimes v_0, I \otimes v_1, Z \otimes v_2, I \otimes v_3, Z
\otimes v_4, Z \otimes v_5, I \otimes v_6,
Z \otimes v_7, \nonumber \\
X \otimes Z \otimes Z \otimes Z, (X \cdot Z) \otimes v_1, X \otimes
v_2, (X \cdot Z) \otimes v_3, X \otimes  v_4, X \otimes  v_5,
(X\cdot Z) v_6, X \otimes  v_7 \}. \label{encoder}
\end{eqnarray}
Due to the decomposition of $\ket{4,2}$ into $\ket{4,2}=
\ket{0}\otimes \ket{3,2}/\sqrt{2}+\ket{1} \otimes
\ket{3,1}/\sqrt{2}$, and using the relations
\begin{eqnarray}
\bra{3,1}v_1\ket{3,2} = -2,\bra{3,2}v_1\ket{3,1} = -2,
\bra{3,1}v_3\ket{3,2} = 2,\nonumber \\
\bra{3,2}v_3\ket{3,1} = 2,
\bra{3,1}v_7\ket{3,2} = 1, \bra{3,2}v_7\ket{3,1} = -1,
\end{eqnarray}
we can directly check the orthogonality of the encoded quantum
states given by Eq.~($\ref{encoder}$).  We still do not have a
construction of local encoders for $\ket{6,2}$ and $\ket{6,4}$
states, thus, it is not proven that they are locally encodable or
not. The summary of locally encodable symmetric basis states and
their encoders are given in Fig.~\ref{dir}.

\section{The Pseudo Clifford set and Pauli encodable states}
\label{SCN: Group analysis}

In the previous section, we have shown a constructive method for finding the local encoder set $\{ v_i
\}$ of symmetric basis states represented by $U \ket{\bar 0}$ where a stronger condition $w_i=v_i$ for
$ v_i  = U w_i U^\dagger$ is satisfied.  However, in general, we can have $\{ w_i \} \neq \{ v_i \}$
(as we have for the more general case in Theorem \ref{theorem:deomposition}). Finding the Pauli sum
representation of a Pseudo Clifford operation $U = \sum_i{a_i p_i}$ compatible with the local encoder
set $\{ v_i\}$ for the general cases is not straightforward. In this section, we consider a reverse
engineering of this problem and present a way to derive a class of locally encodable states for a given
Pauli group local encoder set $\{ v_i\}$ by a group theoretical analysis.

We first investigate general group theoretical properties of a unitary operator $U$  satisfying $U
\sigma U^\dagger = \sigma$ where $\sigma$ is an element of a subset of the Pauli group operations and
derive a formal definition of the Pseudo Clifford set. We start by defining a general subset of unitary
operations.
\begin{definition} \label{subset}
For a subset ${\cal S}$ of the Pauli group ${\cal P}$, the set
${\cal SU_S}$ of $n$-qubit unitary operations is defined by
\begin{eqnarray}
{\cal SU_S} := \{ U\in SU(2^n)\ |\ U s U^{\dagger} \subset {\cal P}
, s \in {\cal S} \}.
\end{eqnarray}
\end{definition}
In this definition, the cardinal number of  ${\cal S}$ is not
necessarily to $2^n$, but some other number $m$.  Therefore, if we
take $m < 2^n$, the set of ${\cal SU_S}$ is a larger set of $U$ than
the case we have investigated for local encoding.

In Definition \ref{subset}, we have defined ${\cal S}$ as a subset
of the Pauli group, however, we can extend ${\cal S}$ to a subgroup
of the Pauli group, since for arbitrary $s_1, s_2 \in {\cal S}$, $ U
s_1 s_2 U^{\dagger} = U s_1 U^{\dagger}U s_2 U^{\dagger} \in {\cal
P}$. Thus, we can choose ${\cal S}$ as a subgroup of the Pauli group
${\cal P}$. Further, we can prove that the cardinal number of ${\cal
S}$ equals that of $U s U^{\dagger}$ as in the following lemma:
\begin{lemma} \label{groupisomorphicmap}
A map $f: s\in {\cal S} \mapsto UsU^{\dagger} \in {\cal
S}^{\prime}\subset SU(2)^{\otimes n}$ is a group isomorphic mapping.
\end{lemma}
\begin{proof}
From the definition, the map $f$ is surjective and homomorphism. For
arbitrary $u_1,u_2 \in {\cal S}^{\prime}$, there exist $s_1, s_2 \in
{\cal S}$ such that $u_1 =U s_1 U^{\dagger}, u_2 = U s_2
U^{\dagger}$ since the map $f$ is surjective. Then, $s_1 = s_2$ if
$u_1 = u_2$.
\end{proof}

From Lemma \ref{groupisomorphicmap}, we give a formal definition of
the Pseudo Clifford set.
\begin{definition} (\textbf{Pseudo Clifford set})\\
For subgroups $P_1$ and $P_2$ ($|{P_1}|=|{P_2}|$) of the Pauli group
${P}$, the corresponding {\it Pseudo Clifford} set
$\mathcal{C}({P_1},{P_2})$ is defined by
\begin{eqnarray}
\mathcal{C}({P_1},{P_2}) := \{ U \in SU(2^n) \ |\ U P_1 U^{\dagger}
= P_2,\ P_1,P_2\subset {\cal P} \}. \label{Def22}
\end{eqnarray}
\end{definition}
\noindent Hence Pseudo Clifford sets are defined with respect to
particular Pauli subgroups $P_1$ and $P_2$ -- and they conjugate one
to the other. The set $\mathcal{C}(P_1, P_2)$ is not necessarily a
group and equals to the Clifford group $\mathcal{C}$ if
$P_1=P_2={\cal P}$.

For local encoding, we require $|P_i|=2^n$. In Theorem
\ref{theorem:deomposition} we choose the set $P_1=\{w_i\}$ which
locally encodes $\ket{\bar{0}}$ (for example
$\{w_i=w^{\bar{0}}_i\}$). From Theorem \ref{theorem:deomposition},
the state $U\ket{\bar{0}}$ can be locally encoded if $U\in
\mathcal{C}(P_1=\{w_i\}, P_2)$, and by definition $P_2$ is the set
of local encoders, given by $P_2=\{ v_i = U w_i U^{\dagger} \}$, as
can easily be seen

\begin{eqnarray}
\bra{\bar 0}  U^{\dagger} v^{\dagger}_i v_j U  \ket{\bar 0} = \bra{\bar 0}  U^{\dagger} v^{\dagger}_i
U^{\dagger} U v_j U \ket{\bar 0} = \bra{\bar{0}} w^{\dagger}_i w_j  \ket{\bar 0}  = \delta_{ij}.
\end{eqnarray}
Our goal in this section is to derive a set of Pauli encodable
states for a given Pauli local encoder $\{ v_i \} \in P_2$.

By denoting a generator of the local encoder set $\{ w_i \} \in P_1$
by $\{ g_i \}_{i=1}^n$, and a generator of the local encoder set $\{
v_i = U w_i U^{\dagger} \} \in P_2$ by $\{ g'_i \}_{i=1}^n$, we
reduce the relationship for the local encoders $\{ w_i\}$ and $\{
v_i\}$ to the relationship for the generators $\{ g_i\}$ and $\{
g'_i\}$. Further, since Clifford operations give a way to map a
generating set of Pauli operations to another generating set, there
always exists a Clifford operation ${C}$ such that
\begin{eqnarray}
C^{\dagger} U g_i U^{\dagger}C = C^{\dagger}g'_i C = g_i, \to Z g_i
Z^{\dagger} = g_i
\end{eqnarray}
for an arbitrary $g_i$, where $Z=C^{\dagger} U$. Therefore, we can
assume that $Z$ maps all the elements of $P_1$ to themselves.  This
result shows that multiplication of appropriate Clifford operation
$C^\dagger$ reduces our problem to a simple case of $g'_i=g_i$ (i.e.
$\{v_i\}=\{w_i\}$), which we have investigated in Section 3.

To obtain a non-Clifford but Pseudo Clifford operation $Z$, we
define an hermitian commutative set ${\cal L}$ by
\begin{eqnarray}
{\cal L}=\{ p \in {\cal P} \ | \ [p,g_i]=0,\ p:\textrm{Hermitian},\
\forall g_i \in \{ g_i \}_{i=1}^n \}. \label{comset}
\end{eqnarray}
Thus, we can construct a subgroup ${\cal SU_L}$ of $SU(2^n)$,
because ${\cal L}$ is a subalgebra of the Lie algebra of $SU(2^n)$.
From the definition of ${\cal L}$, all the elements of ${\cal SU_L}$
commute with all the elements of ${\cal S}$. Therefore, we obtain a
Pauli encodable state given by $C Z \ket{\bar 0} (C\in \mathcal{C},\
Z \in {\cal SU_L})$.

Our method for obtaining the Pauli encoders is summarized in the
following steps.
\begin{enumerate}
\item{Choose a set of Pauli generators $\{ g_i \}_{i=1}^n$ for the zero state $\ket{\bar 0}$.}
\item{Construct a hermitian commutative set ${\cal L}$ defined by Eq.~(\ref{comset}) associated with
the generators $\{ g_i \}_{i=1}^n$.}
\item{Construct a Lie group ${\cal SU_L}$ from the Lie subalgebra ${\cal L}$.
We call the Lie group ${\cal SU_L}$ a Pauli encoder group.}
\item{For an arbitrary Clifford operation $C$ and an arbitrary $Z \in {\cal SU_L}$, we obtain a Pauli
encodable state $C Z \ket{\bar 0}$ and a Pauli encoder  $\langle \{
C g_i C^{\dagger} \}_{i=1}^n \rangle$.}
\end{enumerate}

In general the difficulties arise in step ii) -- not all generator
sets will allow for a nice construction of $\cal L$ (\ref{comset}).
We show two concrete constructions for the Pauli encodable states
for given Pauli encoders in the followings. If the generator set is
given by $\{ g_i = I^{\otimes (i-1)}\otimes X \otimes I^{\otimes
(n-i)} \}$, we have the hermitian commutative set ${\cal L}$ defined
by
\begin{eqnarray}
{\cal L}=\{w_{i}^{\bar{0}} \equiv X^{i_1}\otimes X^{i_2} \otimes
\cdots \otimes X^{i_n},\ i:=i_1i_2\cdots i_n \in \mathbb{Z}_2^{n}
\},
\end{eqnarray}
as used earlier. Thus, an element $Z \in SU_L$ is given by
\begin{eqnarray}
Z = \exp [i\sum_{i\in \mathbb{Z}_2^n}c_i w_i^{\bar{0}}],\ (c_i \in
\mathbb{C}).
\end{eqnarray}
Therefore, with an arbitrary Clifford operation $C$, a Pauli
encodable state is given by
\begin{eqnarray}
C \exp [i\sum_{i\in \mathbb{Z}_2^n}c_i w_i^{\bar{0}}] \ket{\bar 0},
\end{eqnarray}
and the corresponding Pauli encoder set is given by
\begin{eqnarray}
\langle \{ C \cdot (I^{\otimes (i-1)}\otimes X \otimes I^{\otimes
(n-i})) \cdot C^{\dagger} \}_{i=1}^n \rangle.
\end{eqnarray}

If the generator is given by $\{ g_i= Z^{\otimes (i-1)}\otimes X
\otimes I^{\otimes (n-i)} \}_{i=1}^n$, we have the hermitian
commutative set ${\cal L}$ defined by
\begin{eqnarray}
{\cal L}=\langle \{ q_{i} \equiv I^{\otimes (i-1)}\otimes X \otimes
Z^{\otimes (n-i)} \}_{i=1}^n \rangle.
\end{eqnarray}
Thus, an element $Z \in SU_L$ is given by
\begin{eqnarray}
Z = \exp [i\sum_{i=0}^{n}c_i q_i] \ket{\bar 0},\ (c_i \in
\mathbb{C}),
\end{eqnarray}
and, with an arbitrary Clifford operation $C$, a Pauli encodable
state is given by
\begin{eqnarray}
C \exp [i\sum_{i=0}^{n}c_i q_i] \ket{\bar 0}  \label{example2}
\end{eqnarray}
where we take $q_0 = I^{\otimes n}$. The corresponding Pauli encoder
is given by
\begin{eqnarray}
\langle \{ C \cdot (Z^{\otimes (i-1)}\otimes X \otimes I^{\otimes
(n-i)}) \cdot C^{\dagger} \}_{i=1}^n \rangle.
\end{eqnarray}
We see that the symmetric basis state represented by $U_{n,1}$ of
Eq.~({\ref{generalW}}) is a special case of Eq.~($\ref{example2}$).

\section{Conclusion and discussion}

In this work we have looked at the possibility of encoding classical
information onto quantum states by local unitary operations. We have
presented explicit encodings for large sets of states including all
stabiliser and various symmetric basis states. We have introduced
the notion of Pseudo Clifford which unifies these states under one
general local encoding method. Finally, by resorting to group
theoretic analysis we have given a method to find large sets of
states with the same local encodings.

Although the methods used for local encoding presented here are not
as general as possible, they do cover many interesting sets of
states. It seems to be a difficult problem to describe the extent to
which states are covered by our encoding strategies. It remains an
open problem wether all states are locally encodable at all.

We may also be interested in different ways of local encoding for other reasons and applications. For
example in dense coding \cite{Bennett92,Bruss04} we wish to encode the information by acting on only a
subset of the parties (the idea being that then by sending that same subset through a quantum channel,
we can communicate more information than the that allowed by the Holevo bound). The local encoding
presented here could not be used for such a protocol. We can then ask can we extend these results to
consider such protocols, or what can our results say about when we can or cannot.

For example, we know that for optimum dense coding, we must have a maximally entangled state between
the senders and receivers. Imagine we try to dense code using the state $|\psi\rangle =
\frac{1}{\sqrt{2^{n-1}}}\sum_{i=0}^{2^{n-1}}U|i\rangle_s \otimes |i\rangle_r \in \mathcal{H}^{\otimes
n} \otimes \mathcal{H}^{\otimes n}$, where $|i\rangle_x$ are the product states of the computational
basis over the senders' and receivers' spaces for subscripts $s$ and $r$ respectively, and $U$ acts on
$S$ only. The Schmidt basis on the side of the senders (subscript $s$) is in general entangled across
the set of senders by unitary $U$. If $U$ is just identity, then we can encode simply using the Pauli
operators \cite{Bruss04}. Surprisingly we can see that the senders can still encode the full basis
locally, independent of $U$. This can be easily shown, since for optimal dense coding, we would require
$\bra{\psi} v_i^{\dagger} \otimes I \cdot v_j \otimes I \ket{\psi} = \delta_{ij}$ for all $i,j$. It is
easy to see that the $U$ drops out and the condition is equivalent to simply $Tr[v_i^{\dagger}v_j]=0$.
This is satisfied for the local Pauli, hence they allow dense coding for these states. Thus maximum
entangled states can be always optimally locally dense coded independent of the Schmidt basis. The same
result can be obtained by using the Choi-Jamiolkowski isomorphism \cite{ChoiJamiolkowski} by bringing
the unitary over to the receiver's side $|\psi\rangle =
\frac{1}{\sqrt{2^{n-1}}}\sum_{i=0}^{2^{n-1}}U|i\rangle_s \otimes V|i\rangle_r =
\frac{1}{\sqrt{2^{n-1}}}\sum_{i=0}^{2^{n-1}}|i\rangle_s \otimes U^{T}V|i\rangle_r$. In this way it is
clear that the standard Pauli approach will work from \cite{Bruss04}.

We can also note that some of the states considered here have mirror
results in local decoding. It is known that the ability to decode
such encoded classical information is bounded by the entanglement
\cite{Hayashi05}, and explicit bounds are given for $W$-states and
large sets of graph states (which, in the case of graph states, can
be made tight \cite{Markham06b}). We can thus compare the amount of
information we can encode to that we can decode $\Delta I_{local} =
I_{local~encodable}-I_{local~decodable}$. For graph states this
gives  $\Delta I_{local}= E(|\psi\rangle) = n/2$ (where $E$ is the
geometric measure of entanglement \cite{Shimony95}). Indeed, from
\cite{Markham06b} for all states where we can locally encode we have
$\Delta I_{local} \geq E(|\psi\rangle)$. This allows us to talk
about a kind of irreversibility of local information - we can encode
much easier than decode locally, and the difference is bounded by
the entanglement. It is interesting to consider what such a quantity
would mean in relation to other tasks such as measurement based
quantum computing, error correction e.t.c.

We see then that there are many open questions remaining, and that
these results may have potential interest in various areas of
quantum information processing and studies of locality. In addition
we may also consider the usefulness of the task directly in
many-party quantum cryptographic scenarios where we have distributed
encoders and decoders. These will be the topics of ongoing study.

\section*{Acknowledgements}
We are very grateful to Akimasa Miyake, Shashank Virmani, Masaki
Owari, Toshio Ohshima and Terry Rudolph for stimulating discussions
on this topic. This work was sponsored by the Asahi Glass Foundation
and the Japan Society for the Promotion of Science.


\label{lastpage}

\end{document}